\pdfoutput=1

 \documentclass[final,1p,times]{elsarticle}
\usepackage{graphicx}
\usepackage{subfigure}
\usepackage{amssymb}
 \usepackage{amsthm}

\usepackage{lineno}

\journal{Journal of Instrumentation}

\begin{document}
	
	\begin{frontmatter}
		
		%% Title, authors and addresses
		
		\title{Description and simulation of physics of Resistive
			Plate Chambers}
		
		%% use the tnoteref command within \title for footnotes;
		%% use the tnotetext command for the associated footnote;
		%% use the fnref command within \author or \address for footnotes;
		%% use the fntext command for the associated footnote;
		%% use the corref command within \author for corresponding author footnotes;
		%% use the cortext command for the associated footnote;
		%% use the ead command for the email address,
		%% and the form \ead[url] for the home page:
		%%
		%% \title{Title\tnoteref{label1}}
		%% \tnotetext[label1]{}
		%% \author{Name\corref{cor1}\fnref{label2}}
		%% \ead{email address}
		%% \ead[url]{home page}
		%% \fntext[label2]{}
		%% \cortext[cor1]{}
		%% \address{Address\fnref{label3}}
		%% \fntext[label3]{}

		%% use optional labels to link authors explicitly to addresses:
		%% \author[label1,label2]{<author name>}
		%% \address[label1]{<address>}
		%% \address[label2]{<address>}
		
		\author{V. Fran\c cais}
		
		\address{Universit\'e Clermont Auvergne, Universit\'e Blaise Pascal, Universit\'e Blaise Pascal, CNRS/IN2P3, LPC,\\ 4 Av. Blaise Pascal, TSA/CS 60026, F-63178 Aubi\`ere, France}
		
		\begin{abstract}
			%% Text of abstract
			Monte-Carlo simulation of physical processes is an important tool for detector development as it allows to predict signal pulse amplitude and timing, time resolution, efficiency \ldots
			Yet despite the fact they are very common, full simulations for RPC-like detector are not widespread and often incomplete. They are often based on mathematical distributions that are not suited for this particular modelisation and over-simplify or neglect some important physical processes.
			
			We describe the main physical processes occurring inside a RPC when a charged particle goes through (ionisation, electron drift and multiplication, signal induction ...) through the Riegler-Lippmann-Veenhof model together with a still-in-development simulation. This is a full, fast and multi-threaded Monte-Carlo modelisation of the main physical processes using existing and well tested libraries and framework (such as the Garfield++ framework and the GNU Scientific Library). It is developed in the hope to be a basic ground for future RPC simulation developments.
		\end{abstract}
		
		
	\end{frontmatter}
	
	%%
	%% Start line numbering here if you want
	%%
	%\linenumbers
	
	%% main text
%\begin{document}
%	\maketitle
%	\flushbottom
	
	\section{Introduction}
	Resistive Plate Chambers (RPC) are gaseous particle detectors widely used in many High Energy Physics experiments as they are affordable and yet efficient and reliable (compared to scintillator-based detectors), both as timing or tracking charged-particle detector.
	
	A RPC is basically made of gaseous mixture contained between two plates of resistive materials (typically Bakelite or glass) where a high-voltage is applied between them, typically from $1$ to $10$ kV. %(see figure~\ref{fig:RPCGeom}). 
	
	A charged particle going through the detector will ionise the gas, freeing one or more electrons. Those freed electrons, under the influence of the electric field, will drift toward the anode and multiply by interactions with gas molecules and finally produce an electronic avalanche. 
	%\begin{figure}[htbp]
	%	\begin{center}\includegraphics[width=.40\textwidth]{geomCalice.pdf}
	%		%keepaspectratio
	%	\end{center}
	%	
	%	\caption{\label{fig:RPCGeom} Basic geometry of a single-gap RPC detector}
	%\end{figure}
	Typically RPCs are operated with a mixture of three gases: a ionising gas ($\sim 95\%$), an UV quencher gas ($\sim 4\%$) which absorbs photons in order to avoid secondary avalanches, an electron quencher gas ($\sim 1\%$) which absorbs a fraction of the electrons to contain the avalanche.
	
	We present results with the mixture used by CALICE SDHCAL~\cite{calice}: TFE $\mathrm{C_2H_2F_4}$ ($93\%$), $\mathrm{CO_2}$ ($5\%$) and $\mathrm{SF_6}$ ($2\%$). We use a common single-gap RPC geometry: the gas-gap is $1.2$ mm wide, the anode and cathode are respectively $0.7$ and $1.1$ mm thick with a dielectric constant of $\epsilon_r = 7$.
	
	\section{ Primary ionisation   }
\begin{figure}[htbp]
		\centering
		\subfigure[]{\includegraphics[width=.4\textwidth]{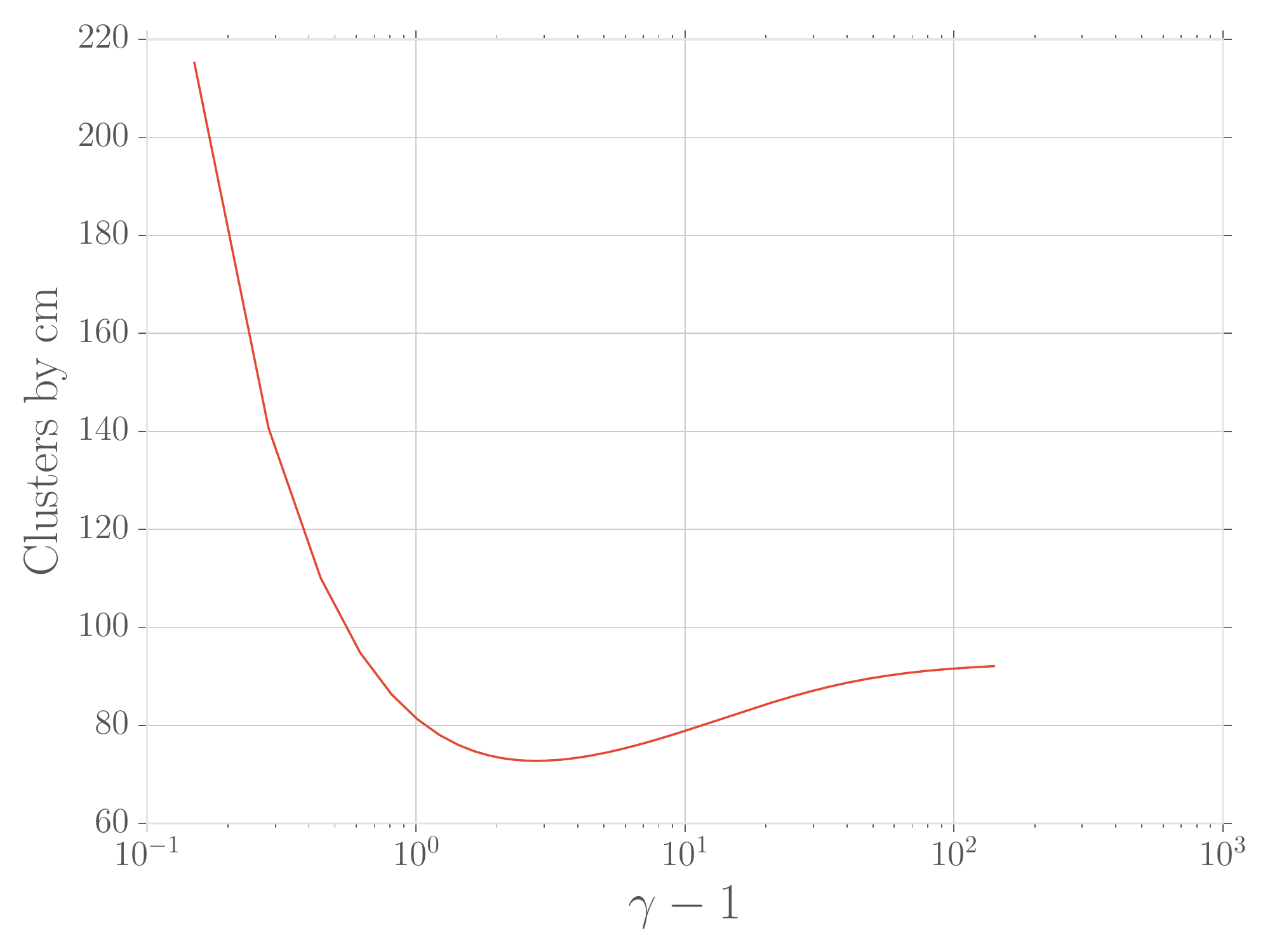} \label{fig:clDens} }
		\qquad
		\subfigure[]{\includegraphics[width=.4\textwidth]{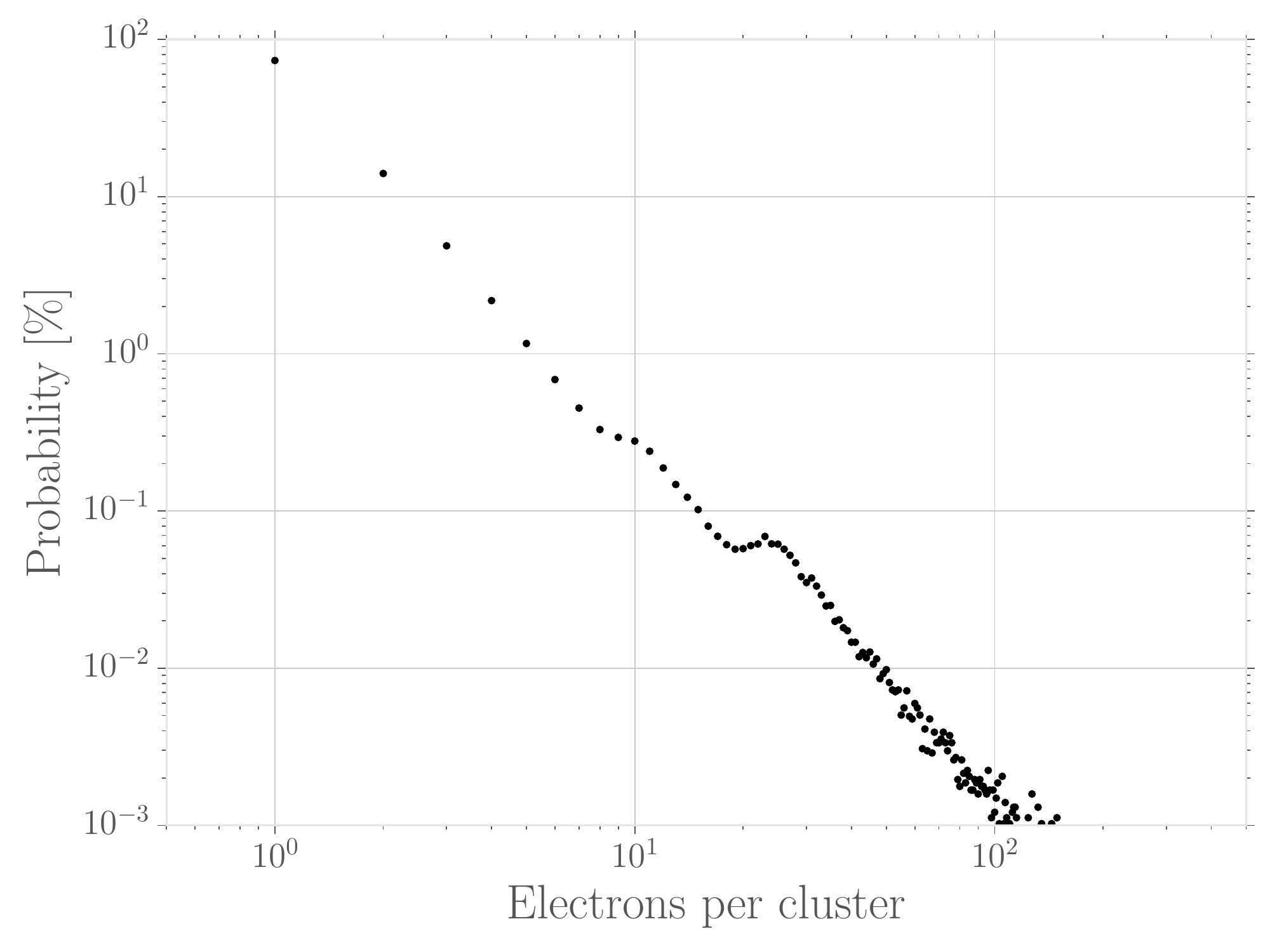} \label{fig:clProba}}
		
		\caption{\subref{fig:clDens} Number of clusters produced by unit of length in function of the particle energy, here a muon. \subref{fig:clProba} Probability distribution for the number of electrons produced per cluster for 4 GeV/c muons.}
	\end{figure}

	When a charged particle traverses the detector it ionises the gas mixture in several zones, by emission of photo-electrons and auto-ionisation (Auger) electrons. Those zones then contain $1$ or more freed electrons which are called electron clusters. The energy deposit is then characterized by the number of clusters produced by unit of length as well as the probability distribution for the number of electrons per cluster. On average a minimum ionising muon produces about $75$ clusters by cm as shown by figure~\ref{fig:clDens}. Most of the time a cluster contains $1$ electron then the probability drops rapidly with the number of electrons and for instance, for $4$ GeV/c muons, is below $1\%$ for more than $5$ electrons, as shown on figure~\ref{fig:clProba}. Those values are computed with the HEED simulation program \cite{Smirnov} and are in good accordance with experimental results  \cite{Smirnov,Riegler}.
	
	%\subsection{Secondary particles impact}
	%Secondary particles produced by the charged particle when interacting with the gas may have a big impact on the avalanche and RPC performance, as they create many ionisation electrons.
	%
	%RPC are often confined by an aluminum case. According to W. Riegler and H. Vincke, a $7$ GeV pion crossing an aluminum plate of $3$ mm has a probability to be accompanied by at least one charged particle of only 4.92\% \cite{Riegler}. Hence secondary particles should not impact RPC performance.
	\section{The Riegler-Lippmann-Veenhof model for electronic avalanche}
	The electrons freed during ionisation will then drift towards the anode under the influence of the electric field, and start an electronic avalanche by multiplying while interacting with gas molecules. In this section we briefly detail the Riegler-Lippmann-Veenhof model for electronic avalanche \cite{Riegler} which is a continuation of the Legler model for avalanche in electro-negative gas~\cite{legler}. %they number will grow by interacting with the gas molecules and start an electronic avalanche.
	\subsection{Electron multiplication}
	The avalanche development is characterised by the Townsend coefficient $\alpha$ and the attachment coefficient $\eta$. If an avalanche contains $n$ electrons at the position $x$, the probability it contains $n+1$ at $x+\mathrm{d}x$ is given by $n\alpha \mathrm{d}x$. In the same way, the probability that one electron gets attached in an avalanche containing $n$ electrons is given by $n\eta \mathrm{d}x$. Then the average numbers of electrons $\overline{n}$ and positive ions $\overline{p}$ are modelised following \cite{Riegler}
	\begin{equation}
		\frac{\mathrm{d} \overline{n}}{\mathrm{d}x} = (\alpha - \eta) \overline{n}, \qquad \frac{\mathrm{d} \overline{p}}{\mathrm{d} x} = \alpha \overline{n}.
	\end{equation}
	With the initial condition $\overline{n}(0)=1$ and $\overline{p}(0)=0$ this yields
	\begin{eqnarray}
		\label{eq:nbar}
		\bar{n}(x) &=& e^{\left(\alpha - \eta \right)x}, \\
		\bar{p}(x) &=& \frac{\alpha}{\alpha - \eta} \left(e^{\left(\alpha - \eta \right)x} - 1 \right).
	\end{eqnarray}
	In order to observe an electronic avalanche the exponential in eq.~\ref{eq:nbar} has to be positive. 
	\begin{figure}[htbp]
		\centering
		\subfigure[]{\includegraphics[width=.49\textwidth]{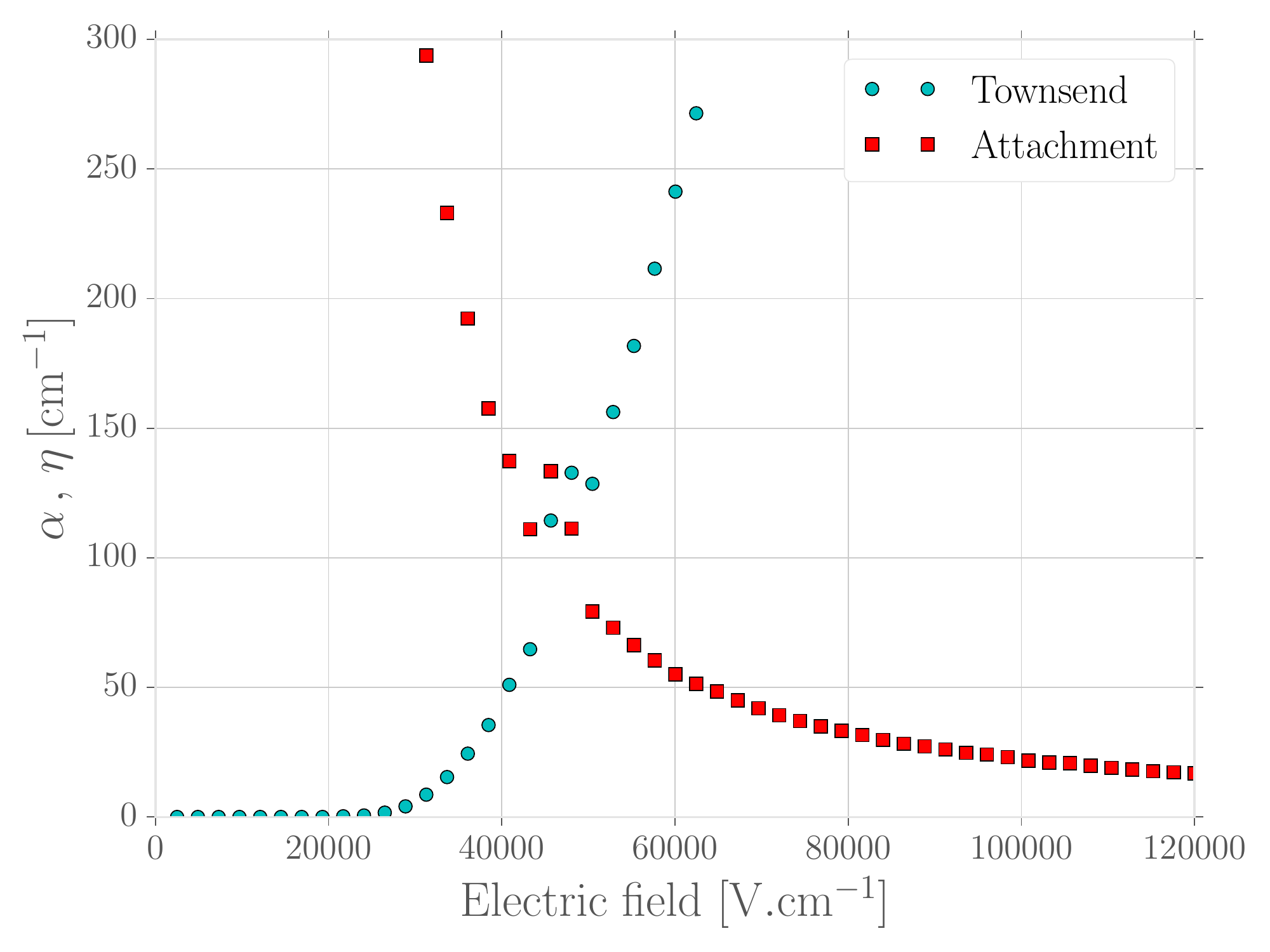} \label{fig:townsend}}
		\subfigure[]{\includegraphics[width=.49\textwidth]{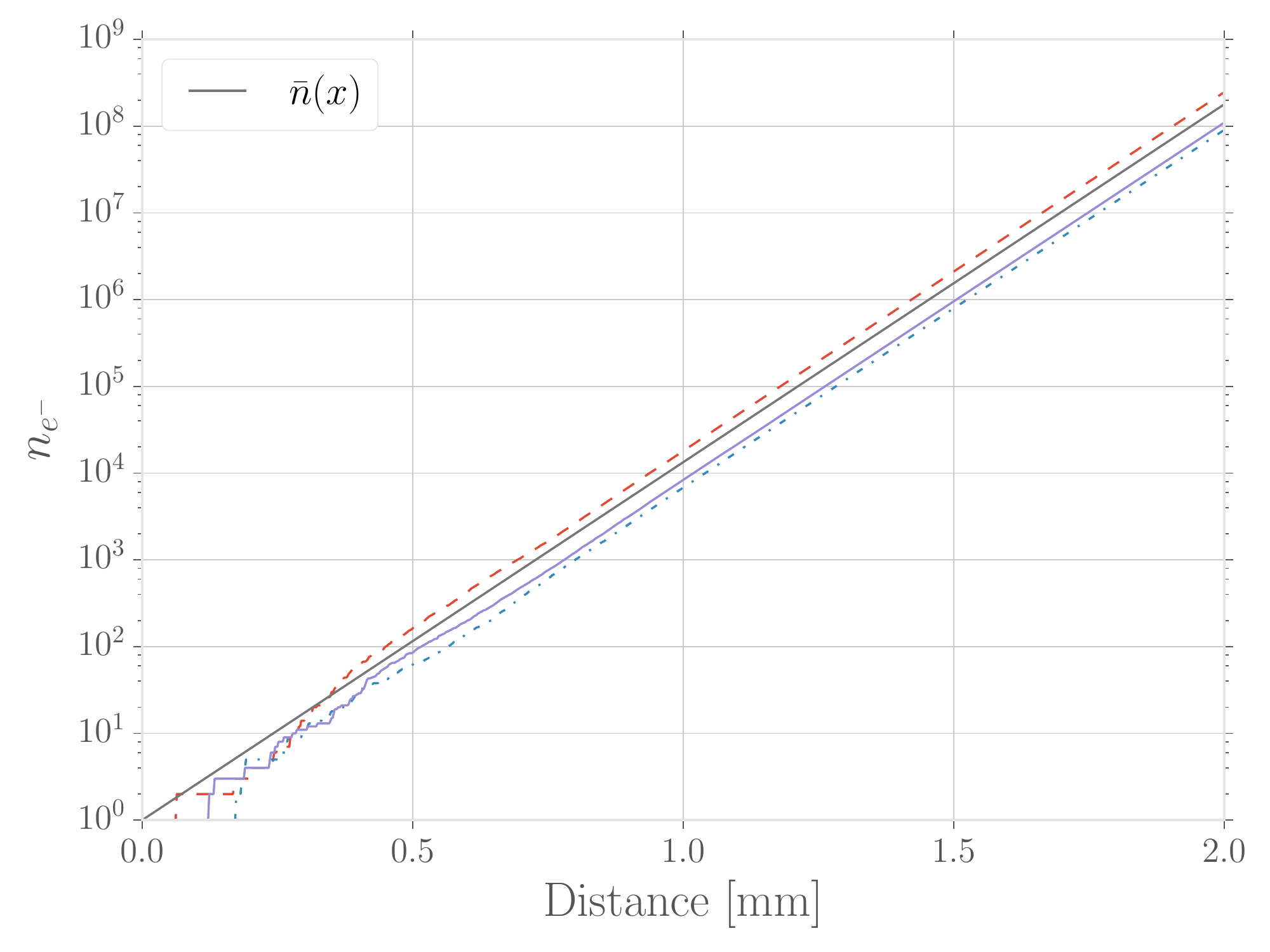} \label{fig:mult}}
		\caption{\subref{fig:townsend}Townsend and attachment coefficients $\alpha$ and $\eta$ as a function of the electric field, for the gaseous mixture of $\mathrm{C_2H_2F_4}$/$\mathrm{CO_2}$/$\mathrm{SF_6}$ ($93\%$/$5\%$/$2\%$). Values are computed with Magboltz 9.01 \cite{Magboltz}. \subref{fig:mult} Simulation of three distinct avalanches started by one electron at $x=0$. The analytic formulation for the average number of electrons (eq.~\ref{eq:nbar}) is also plotted on the figure.}
	\end{figure}
	Figure~\ref{fig:townsend} shows both coefficients in function of the intensity of the electric field. In this case, the electric field needs to be greater than $49$ kV/cm for an avalanche to develop.
	
	The evolution of the number of electrons in an avalanche is modelised following \cite{Riegler}
	\begin{equation}
		\label{eq:n}
		n = \left\{
		\begin{array}{l c r}
			0,& s<k\frac{\bar{n}(x)-1}{\bar{n}(x)-k}  \\
			1 +  \ln \left( \frac{\left(\bar{n}(x)-k\right)\left(1-s\right)}{\bar{n}(x)\left(1-k\right)} \right) \frac{1}{\ln \left(1 - \frac{1-k}{\bar{n}(x)-k}\right)},& s>k\frac{\bar{n}(x)-1}{\bar{n}(x)-k} 
		\end{array}
		\right.
	\end{equation}
	where $s$ is a random number $\in [0,1)$ and $k=\eta/\alpha$. Eq.~\ref{eq:n} is valid only for $\alpha,\eta > 0$, details for the other cases are in \cite{Riegler}. In order to calculate the induced signal we can't use the probability distribution to approximate the final avalanche charge, instead we have to simulate the actual avalanche development. The gas gap is divided in $N$ steps of $\Delta x$. In the case we have $n_0$ electron at the position $x_0$, we'll find $n_1$ electrons at the position $x=x_0+\Delta x$ according to eq.~\ref{eq:n}. Meaning that we loop over the $n_0$ electrons, draw a number from eq.~\ref{eq:n} and sum them. In the same way the $n_1$ electrons will multiply and we'll have $n_2$ electrons at $x=x_0+2\Delta x$. This procedure is iterated until all the electrons have reached the anode.
	
	The figure~\ref{fig:mult} shows the simulation of three avalanches started by one electron at the cathode. The first interactions at beginning of the avalanches has an important influence for their development, and they quickly behave like an positive exponential just like the average number of electrons $\overline{n}(x)$ given by eq.~\ref{eq:nbar}. 
	
	It is also important to note that an avalanche is not only driven by $\left(\alpha - \eta\right)$, but also by $\alpha$ and $\eta$ themselves as pointed out in~\cite{Riegler}.
	\subsection{Diffusion}
	When no electric field is present, an electron cloud in a gas is subject to the classic thermal diffusion. But when an electric field is applied the thermal diffusion motion is superposed by a drift motion. At the microscopic scale an electron drifting the distance $\delta z$ gains the energy $e_0\,|\vec{E}|\,\delta z$ between two collisions, where $|\vec{E}|$ is the intensity of the field and $e_0$ the unit charge. Some of this energy will be lost during the next encounter with a gas molecule (non-ionising elastic collision). Then the electron is again accelerated by the electron field and lost some when colliding, and so on. At a macroscopic scale one sees the electron moving with a constant drift velocity which depends on the gas and the electric field applied \cite{Lippman}. 
	
\begin{figure}[htbp]
		\centering
		\subfigure[]{\includegraphics[width=.49\textwidth]{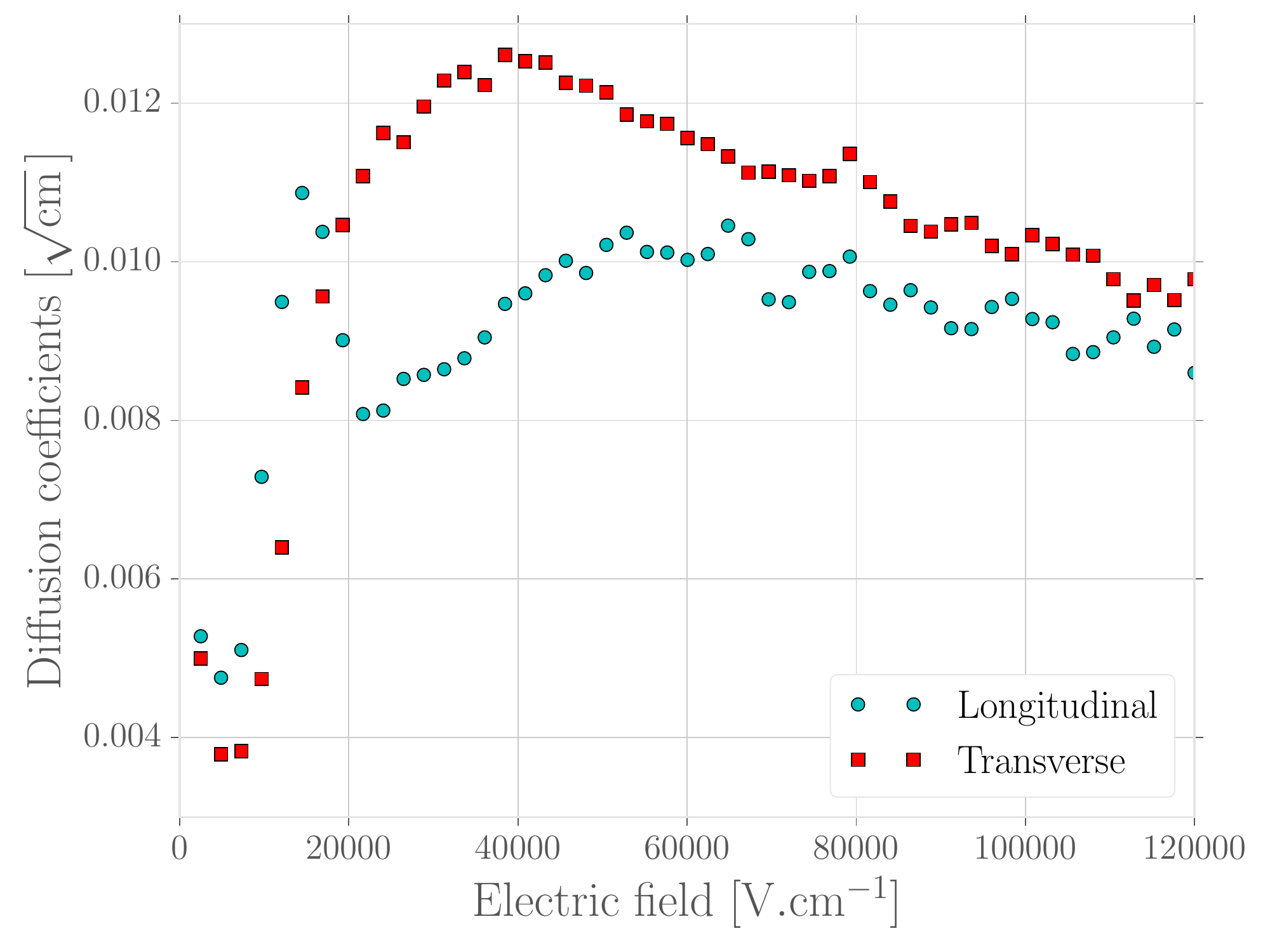} \label{fig:diff} }
		\subfigure[]{\includegraphics[width=.49\textwidth]{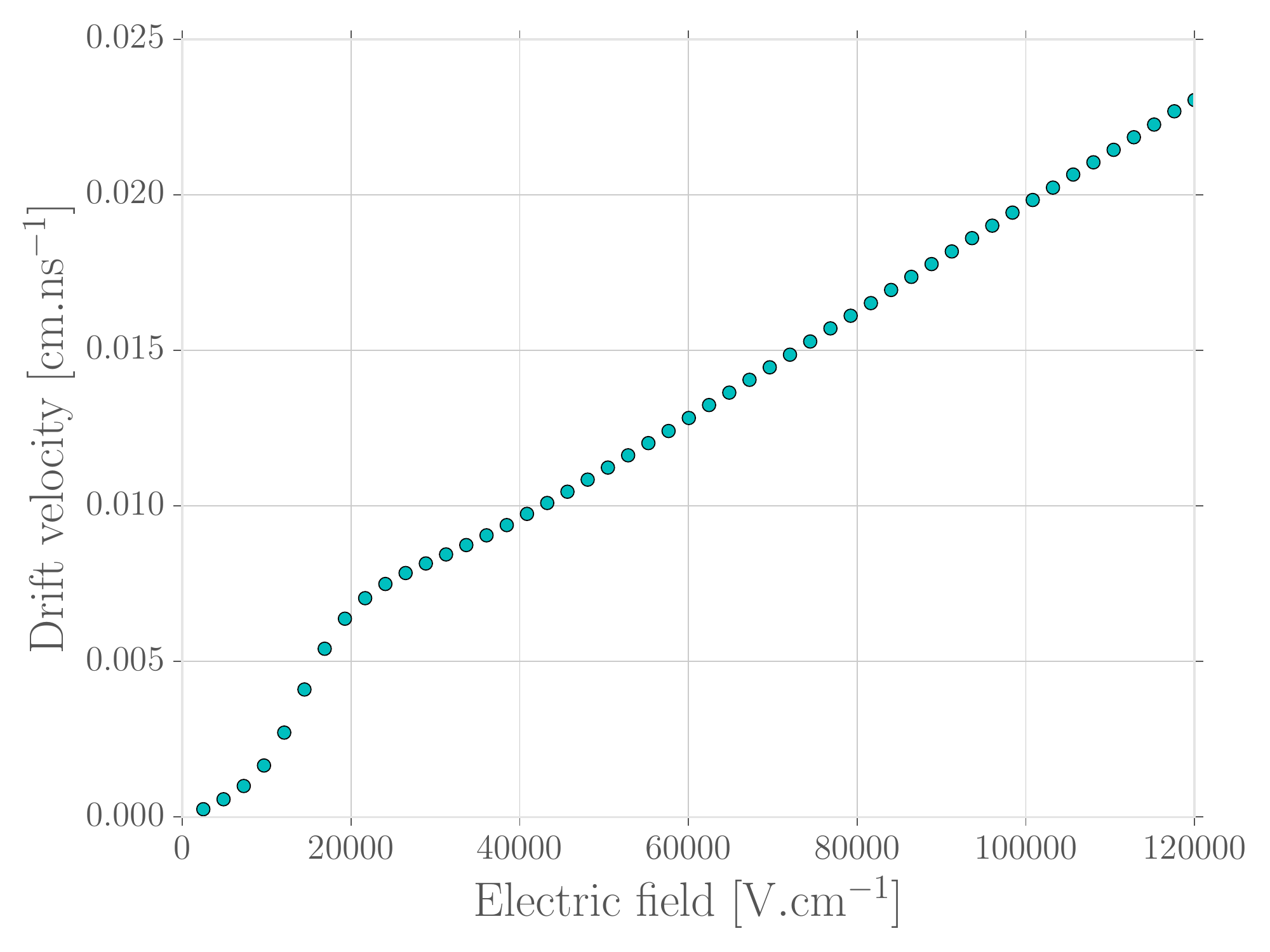} \label{fig:driftV}}
		
		\caption{\subref{fig:diff} Longitudinal and transverse diffusion coefficient as a function of the electric field. \subref{fig:driftV} Electron drift velocity as a function of the field intensity. Both are computed with MAGBOLTZ 9.01 \cite{Magboltz}}
	\end{figure}

	Because of this constant drift motion superposed to the thermal diffusion motion, the diffusion becomes anisotropic. We can separate it in two distinct terms, a longitudinal one and a transverse one~\cite{Lippman}
	\begin{eqnarray}
		\label{eq:diffL}
		\varphi_L(z,l) &=& \frac{1}{\sqrt{2\pi} \sigma_L } \exp{\left( -\frac{\left(z-z_0\right)^2}{2\sigma_L^2} \right)} \\
		\label{eq:diffT}
		\varphi_T(z,l) &=& \frac{1}{\sigma_T^2} \exp{\left( - \frac{\left( r-r_0 \right)^2}{2\sigma_T^2} \right)}
	\end{eqnarray}
	We used cylindrical coordinate and rotational symmetry (a $\phi$-integration was carried out leading to an additional factor of $2\pi$). The width of the Gaussian depends on the longitudinal and transverse diffusion coefficients $D_{L,T}$ $\left[\sqrt{\mathrm{cm}}\,\right]$ and the drifted distance $l$: $\sigma_{L,T} = D_{L,T}\sqrt{l}$. Figure~\ref{fig:diff} shows the dependence of the diffusion coefficient on the electric field.
	
	In the simulation the longitudinal diffusion is fully modelised: the electrons are redistributed at each simulation step where the new $z$-coordinate is computed by drawing a random number from eq.~\ref{eq:diffL}, where $\sigma_L = D_L \sqrt{\Delta z}$ and $\Delta z$ is the detector step which is the distance drifted by the electrons between two simulation steps. However the transverse diffusion is only approximate as the avalanche is modelised in $1$-dimension. We suppose the electrons are contained in a disc perpendicular to the $z$-axis with a radial charge distribution following eq.~\ref{eq:diffT} with $\sigma_t = D_T \sqrt{l}$ with $l$ the distance drifted by the electrons from their position of generation. 
	\subsection{Space Charge Effect}
	This section briefly details the Space Charge Effect in an RPC. The general solution is detailed in~\cite{SCE2} and a description of this effect in RPC can be found in~\cite{SCE,Lippman}.
	
	When the number of charges in the avalanche becomes high enough they influence the electric field, and so influence the values of the Townsend and attachment coefficient $\alpha$ and $\eta$. This is the Space Charge Effect. We need to compute the contribution to the electric field of all the charges present in the avalanche. The analytic formula for the potential $\Phi(r,\phi,z,r',\phi',z')$ of a point charge in an infinite plane condenser of three layers is detailed in \cite{SCE,SCE2}. $(r,\phi,z)$ is the point of observation and $(r',\phi',z')$ is the position of the charge. Since the simulation is in $1$-dimension it is sufficient to compute $\Phi(r=0,\phi=0,z,r',\phi'=0,z')$ = $\Phi(z,r',z')$. As we only approximate the transverse diffusion where we consider a unit charge at position $z'$ is contained in a disk perpendicular to the $z$-axis, the electric field of this disk is given by~\cite{Lippman,SCE}
	\begin{equation}
		\label{eq:Ebar}
		\overline{E}(z,l,z') = - \int_{0}^{\infty} \varphi_T(r',l) \; \frac{\partial \Phi(z,r',z')}{\partial z} \; r' dr
	\end{equation}
	where $\varphi_T$ is the radial charge distribution of the disk (eq.~\ref{eq:diffT}).
	The total field from the space charges is then computed by summing over the disk at each detector bin:
	\begin{equation}
		\label{eq:SCField}
		E_{SC} (z) = \sum_{n=0}^{N} q_n \overline{E}(z_n,l_n,z_n')
	\end{equation}
	where $q_n$ is total charge in bin $n$.
	\subsection{Induced current}
	%Electrons in the gas are not collected by the electrodes as they are absorbed by the resistive layer upstream. In consequence it is the electrons movement in the electric field that induces charges on the electrode. 
	In order to compute the induced current we make use of the Ramo's theorem generalised to resistive materials \cite{Ramo,Ramo2}
	\begin{equation}
		\label{eq:ramo}
		i(t) = e_0\,N(t)\, v_e\, \frac{\varepsilon_r}{\left(d_{r_1}+d_{r_2}\right)+\varepsilon_r d_g} 
	\end{equation}
	where $N(t)$ is the number of electrons in the detector at time $t$, $\epsilon_r$ is the relative dielectric constant of the resistive layers, $d_g$ and $d_{r_i}$ are respectively the width of the gas gap and of the resistive layers. $v_e$ is the electron drift velocity, it depends on the gas mixture as well as the electric field. It is shown on figure~\ref{fig:driftV} as a function of the field intensity.
	\section{Results}
\begin{figure}[htbp]
		\centering
		\subfigure[]{\includegraphics[width=.49\textwidth]{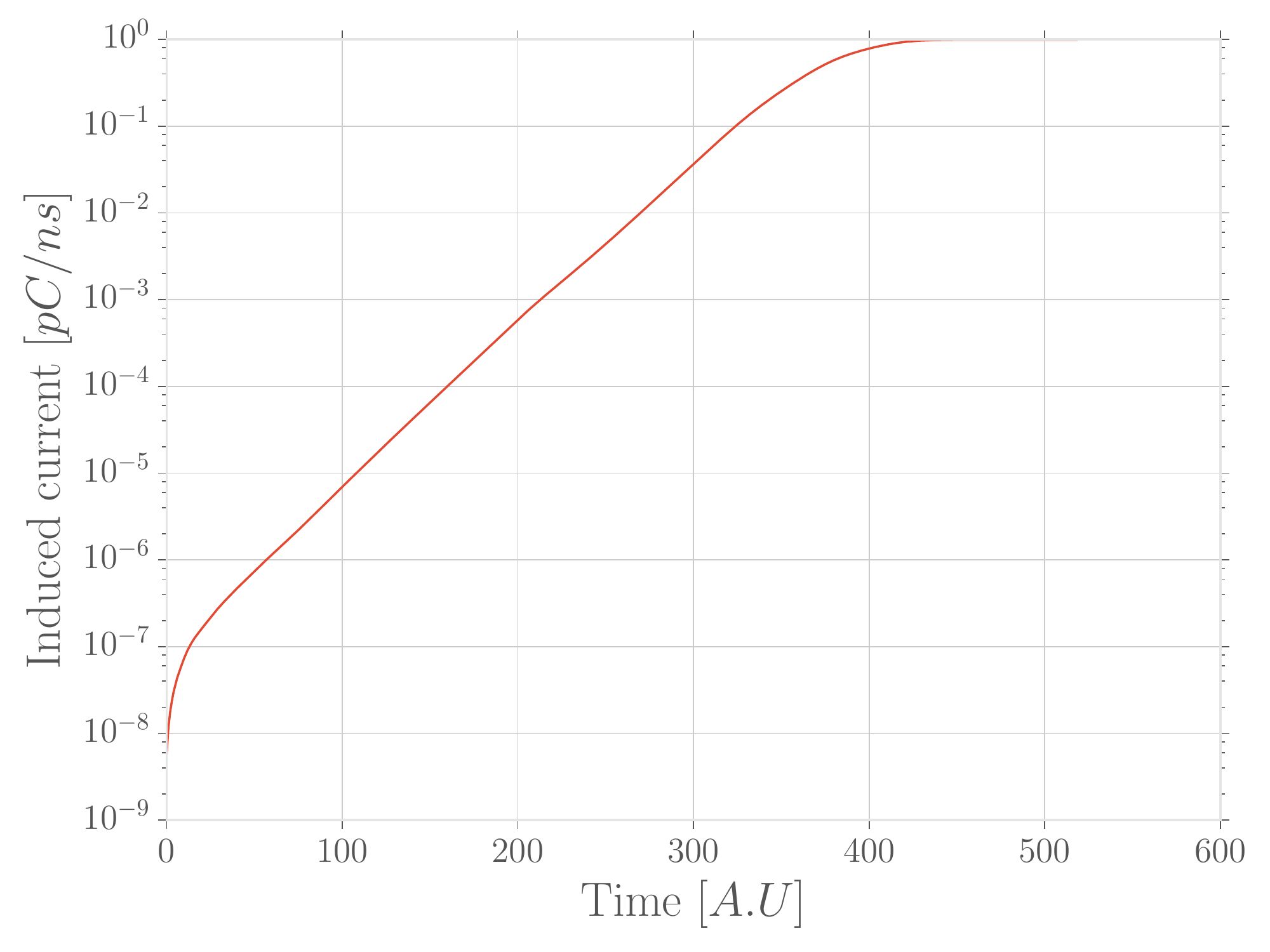}\label{fig:inducedCharges}}
		\subfigure[]{\includegraphics[width=.49\textwidth]{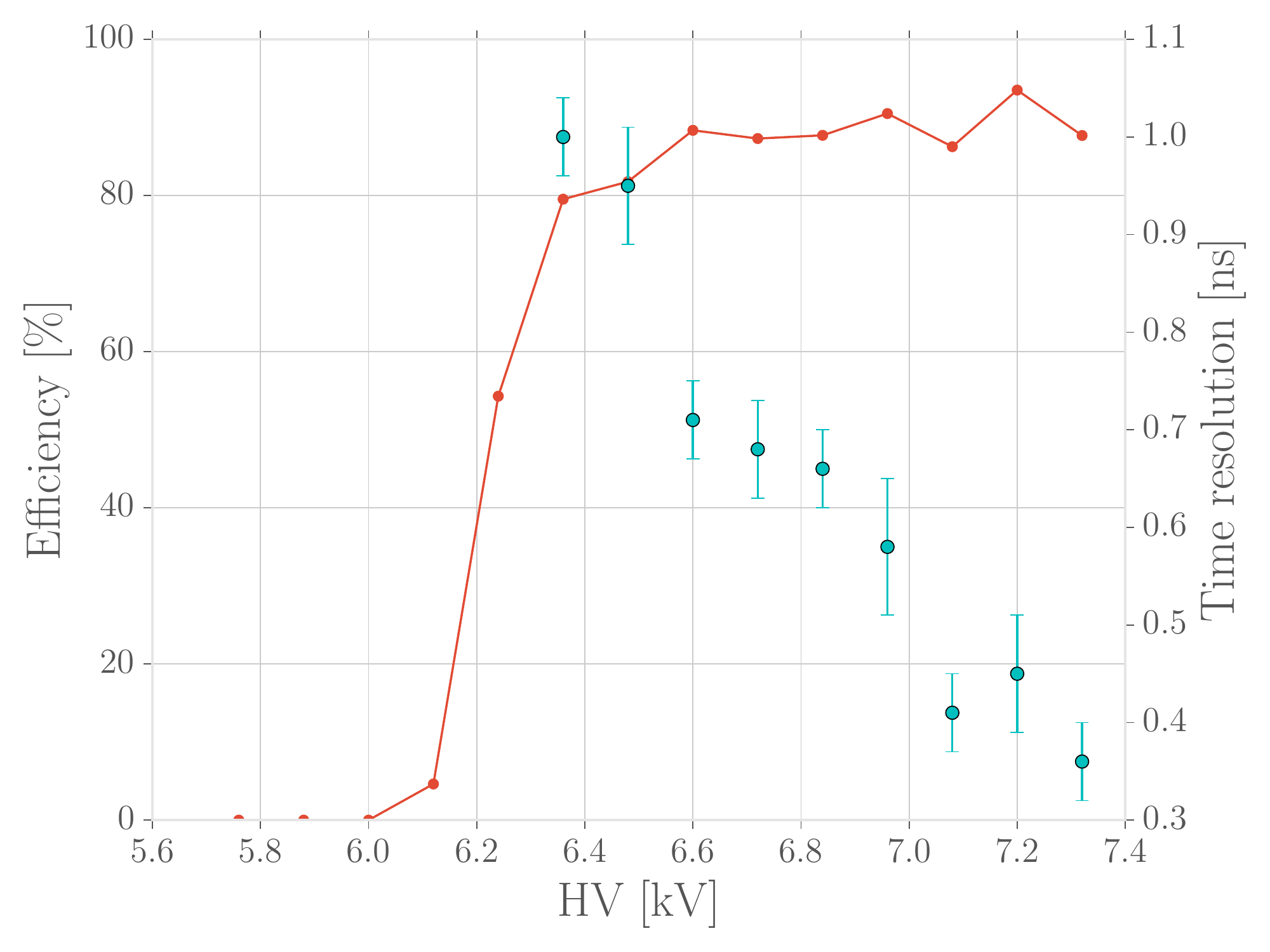}\label{fig:eff}}
		
		\caption{\subref{fig:inducedCharges} Induced charges during an avalanche as function of time (arbitrary unit) for $5$ GeV/c muons. \subref{fig:eff} Simulated efficiency and time-resolution for $5$ GeV/c muons.}
	\end{figure}

	%As the simulation is still work in progress, what is presented here is more a \emph{proof of concept} than actual results, proof that the simulation gives coherent results but still needs to be compared to experimental data.
	The charges induced over time is shown on figure~\ref{fig:inducedCharges}. One can note a clear saturation effect at about time step $350$, due to the Space Charge Effect as the number of charges became high enough to lower the applied field and thus the multiplication gain. Figure~\ref{fig:eff} is a simulated efficiency curve along with the time-resolution. The efficiency is the ratio of the number of avalanches that has crossed the detection threshold (set to $100$ fC) to the total number of avalanches simulated. The time-resolution is taken as the standard-deviation from the threshold-crossing time distribution, i.e. the time when the induced charge has crossed the detection threshold. For the RPC configuration we considered in the simulation, at its operation point of $6.9$ kV, the experimental efficiency is about $96\%$ with a typical time resolution around $500$ ps~\cite{calice} which are in good accordance with figure~\ref{fig:eff}. %(large uncertainties comes from a lack of statistics because of simulation crashes).
	%For this particular detector setup, at his operation point of $6.9$ kV, the experimental efficiency is about $93\%$ and a typical 
	
	\section{Conclusion}
	We have briefly detailed the main physical processes having an important impact for the electronic avalanches occurring in a RPC, through the Riegler-Lippmann-Veenhof model. The results produced by the simulation show good behavior but could be compared to experimental data from test beam. In this so-called 1.5D model, the transverse diffusion is only approximated while a full $2$-dimensional simulation would be, rationally, more accurate but also much slower. This model calls for a lot of matrix algebra, thus taking advantage of the computing power of GPUs for this matter could provide a non-negligible speed-up. 
	%Indeed the main bottle-neck comes from the generation of the large quantity of pseudo-random numbers needed by the simulation. In this case using the computing power of GPUs to generate the pseudo-random numbers (with library such as Nvidia CuRand or Thrust) could provide significant speed-up and allowing the computation of a $2$D model in acceptable time.

	%\acknowledgments
	
	%This is the most common positions for acknowledgments. A macro is
	%available to maintain the same layout and spelling of the heading.
	
	% We suggest to always provide author, title and journal data:
	% in short all the informations that clearly identify a document.

\end{document}